\documentclass[11pt]{article}
\usepackage{moriond,psfig,epsfig}

\def\Journal#1#2#3#4{{#1} {\bf #2}, #3 (#4)}

\def\aap{A\&A}
\def\apj{ApJ}

\def\apjs{APJS}
\def\aj{AJ}

\def\be{\begin{equation}}
\def\ee{\end{equation}}
\def\bea{\begin{eqnarray}}
\def\eea{\end{eqnarray}}

\begin{document}
\vspace*{4cm}
\title{FOSSIL RADIO PLASMA IN CLUSTER MERGER SHOCK WAVES}
\author{T.A. EN{\SS}LIN$^1$, M. BR{\"U}GGEN$^{1,2}$}
\address{$^1$ Max-Planck-Institut f{\"u}r Astrophysik, Garching D-85741,
	Germany\\
	$^2$ Institute of Astronomy, Madingley Road, Cambridge CB3 0HA, United Kingdom}
\maketitle
\abstracts{In several merging clusters of galaxies so-called {\it
cluster radio relics} have been observed. These are extended radio
sources which do not seem to be associated with any radio galaxy. Two
competing physical mechanisms to accelerate the radio emitting
electrons have been proposed: (i) diffusive shock acceleration and
(ii) adiabatic compression of fossil radio plasma by merger shock
waves.  Here the second scenario is investigated.  We present detailed
3-dimensional magneto-hydrodynamical simulations of the passage of a
radio plasma cocoon through a shock wave. Taking into account
synchrotron, inverse Compton and adiabatic energy losses and gains we
evolved the relativistic electron population to produce synthetic
radio maps in Stokes I-, Q-, and U-polarisation. In the synthetic
radio maps the electric polarisation vectors are mostly perpendicular
to the filamentary radio structures.}

\section{Introduction}

In the current picture of hierarchical structure formation clusters of
galaxies grow mainly by the merging of smaller and moderately sized
sub-clusters. During such merger events a significant fraction of the
kinetic energy of the intergalactic medium (IGM) is dissipated in
Mpc-sized shock waves. The shock waves are responsible for heating the
IGM to temperatures of several keV. Moreover, they are likely to inject
and accelerate relativistic particle populations on cluster scales.

Cluster-wide relativistic electron populations are indeed observed in
several merging or post-merging clusters. The so-called {\it cluster
radio relics} are are believed to be directly related to cluster
mergers (En{\ss}lin et al. 1998, Roettiger et al. 1999, Venturi et al.
1999, En{\ss}lin \& Gopal-Krishna
2001)\nocite{1998AA...332..395E,1999ApJ...518..603R,1999dtrp.conf..27V,2001A&A...366...26E}. Like
the radio halos they are extended radio sources with a steep
spectrum. In the literature radio relics are often confused with radio
halos even though several distinctive properties exist. Cluster radio
relics are typically located near the periphery of the cluster; they
often exhibit sharp emission edges and many of them show strong radio
polarisation.

In several cases it could be shown that shock waves are present at the
locations of the relics. Moreover, the cluster radio relic 1253+275 in
the Coma cluster shows a morphological connection to the nearby radio
galaxy NGC 4789 (Giovannini, Feretti \& Stanghellini
1991\nocite{1991A&A...252..528G}). This suggests that radio relics may
be fossil radio plasma that has been revived by a shock. Fossil radio
plasma is the former outflow of a radio galaxy in which the
high-energy radio emitting electrons have lost their energy. Due to
their invisibility in the radio these cocoons are also called {\it
radio ghosts} (En{\ss}lin 1999)\nocite{1999dtrp.conf..275E}.

The first relic formation models considered diffusive
shock acceleration (Fermi I) as the process producing the radio
emitting electrons (En{\ss}lin et al. 1998, Roettiger et al. 1999,
Venturi et al.
1999)\nocite{1998AA...332..395E,1999ApJ...518..603R,1999dtrp.conf..27V}.
However, when a fossil radio cocoon is passed by a cluster merger
shock wave, with a typical velocity of a few 1000 km/s, the cocoon is
compressed adiabatically and not shocked, owing to the much higher
sound speed within it. Therefore, shock acceleration cannot be the
mechanism that re-energises the relativistic electron population in
the cocoon. But the energy gained during the adiabatic compression
combined with the increase in the magnetic fields strength can make a
fossil radio cocoon emit radio waves again. One prerequisite for this
is that the electron population is not older than 0.2 - 2 Gyr
(En{\ss}lin \& Gopal-Krishna 2001\nocite{2001A&A...366...26E}).

En{\ss}lin \& Gopal-Krishna (2001)\nocite{2001A&A...366...26E} showed
that the spectral properties of cluster radio relics are well
reproduced by this scenario. Here and in En{\ss}lin \& Br\"uggen (2001),
we demonstrate that the observed morphologies and polarisation
patterns are reproduced by this model as well. This is done with the
help of the first 3-dimensional magneto-hydrodynamical (MHD)
simulations of a fossil radio cocoon that is passed by a shock
wave. We produce artifical radio maps that can be compared directly to
new high-resolution radio maps of cluster radio relics.

\begin{figure*}
\begin{center}
\psfig{figure=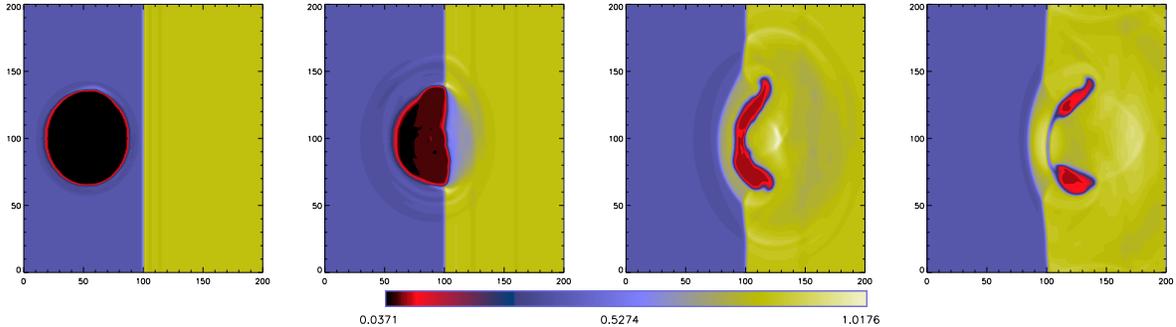,width=1.0 \textwidth,angle=0}
\end{center}
\caption[]{\label{fig:evolution1} Evolution of the gas density in a central slice through
the simulation volume. The formation of a torus is visible. }
\end{figure*}

\section{Method}
\subsection{3D MHD Simulation}

The magneto-hydrodynamical simulations were obtained using the ZEUS-3D
code which was developed especially for problems in astrophysical
hydrodynamics (Stone \& Norman 1992a,
b)\nocite{1992ApJS...80..753S,1992ApJS...80..791S}.  

The simulations were computed on a Cartesian grid with 100$^3$ and
200$^3$ equally spaced zones. In the units of the simulation the
computational domain ranged form 0 to 10 in each coordinate. The
simulation was set up such that a stationary shock formed at $x=5$
that is perpendicular to the direction of the flow.

In the pre-shock region a spherical bubble was set up, in
which the density was lowered by a factor of 10 with respect to the
environment. In turn, the temperature in the bubble was raised such
that the bubble remained in pressure equilibrium with its
surroundings. Inside the bubble a magnetic field was set up which was
computed from a random field that consisted of several hundred Fourier
modes. Finally, the bubble was filled with around $10^4$ uniformly
distributed tracer particles that are advected with the flow.

\subsection{Radio Maps}

The radio maps are constructed using the tracer particles. Initially,
each tracer particle is located inside the radio plasma cocoon and is
associated with the same initial relativistic electron
population. Then the electron spectrum for each tracer particle is
evolved in time taking into account synchrotron, inverse Compton, and
adiabatic energy losses and gains (see En{\ss}lin \& Gopal-Krishna
2001\nocite{2001A&A...366...26E}).

The pre-shock external gas density is set to about $5\cdot
10^{-4}\,{\rm electrons/cm}^3$ and the temperature to 1-2 keV. The
simulation box is assumed to have a size of 1 Mpc$^3$ and to be
located at a distance of 100 Mpc from the observer.

\begin{figure}
\begin{center}
\psfig{figure=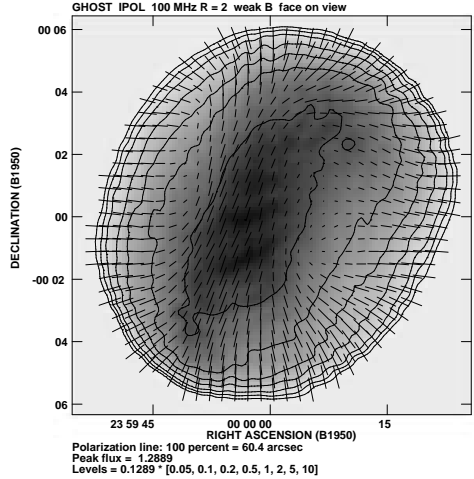,width=0.40 \textwidth,angle=0}
\end{center}
\caption[]{\label{fig:map.WW+10-00} Radio emission of the radio bubble
before shock passage. The shock has a compression factor of $C=2$.
The polarisation E-vectors are displayed by dashes with the length
proportional to the relative polarisation. Here and in the following
radio maps the flux is given in Jansky per simulation pixel (with
linear size of 20.6 arcsec)}
\end{figure}

\begin{figure}
\begin{center}
\psfig{figure=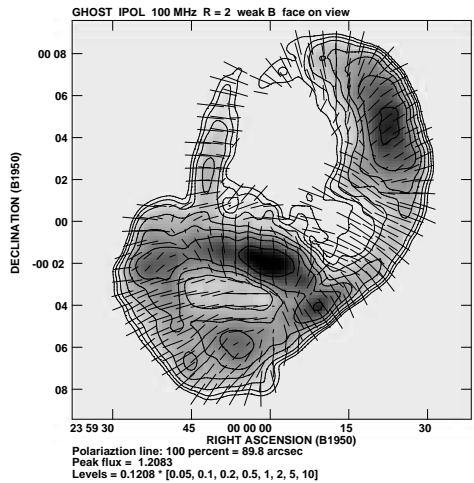,width=0.40 \textwidth,angle=0}
\end{center}
\caption[]{\label{fig:map.WW+70-00} Same as Fig. \ref{fig:map.WW+10-00}. Late
stage of the shock passage in the same model.}
\end{figure}

\begin{figure}
\begin{center}
\psfig{figure=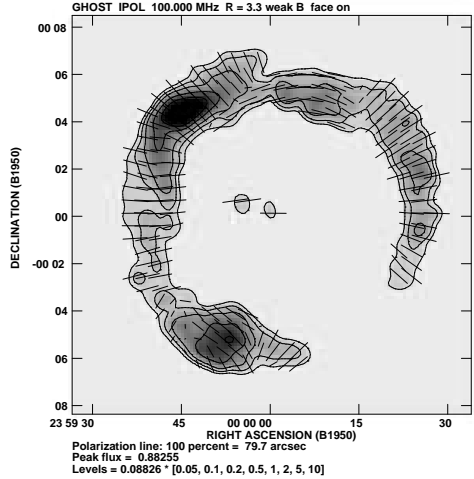,width=0.40 \textwidth,angle=0}
\end{center}
\caption[]{\label{fig:map.SW70-00} Edge-on view of the late stage of
the shock passage in a model, where the shock has a compression factor
of $C=3.3$. Note the reversal of the polarization vector
orientaion. For details see Fig. \ref{fig:map.WW+10-00}.}
\end{figure}

An increase in the radio luminosity during the compression can be seen
in the simulation.  However, the strong increase expected from the
analytical model of En{\ss}lin \& Gopal-Krishna
(2001)\nocite{2001A&A...366...26E} could not be reproduced here. This
is a result of the decaying magnetic fields due to numerical
resistivity, and not a failure of the model. In order not to be too
much affected by the rapidly evolving spectral cutoff we restrict our
analysis in the following to a low radio frequency, namely 100 MHz.

\subsection{Radio Morphology}

As the shock wave passes the initially spherical radio cocoon (Fig.
\ref{fig:map.WW+10-00}), the cocoon is torn into a filamentary
structure (Fig.  \ref{fig:map.WW+70-00}). In the simulations with weak
magnetic fields the final morphology is toroidal for a strong shock,
Fig. \ref{fig:map.SW70-00}, and shows two tori in the case of a weak
shock wave, Fig. \ref{fig:map.WW+70-00}). Such a double torus seems to
be observed in the relic in Abell 85 (Slee et
al. 2001\nocite{slee2001}). The degree of polarisation is relatively
high everywhere. The electric polarisation vectors tend to be
perpendicular to the radio filaments indicating aligned magnetic field
structures within them. At some few spots the polarisation E-vectors
are aligned with the radio filaments (see Fig. \ref{fig:map.SW70-00}).

\subsection{Radio Polarisation}

The total integrated polarisation flux of a radio relic observed
face-on can be expected to be small. In the case of a toroidal relic
all polarisation orientations are roughly equally present and
therefore cancel each other out in the surface integration. But even
in a more complex relic the polarisation cancels out in the face-on
view.

In the edge-on view this is different. Now a preferential direction
exists: E-vectors tend to be aligned with the projected shock normal.
Whenever the imprinted polarisation dominates over the initial
intrinsic one, the direction of the E-vector can be used relatively
reliably to infer the shock normal projected onto the plane of the
sky. The angle with respect to this plane can be estimated roughly
from the total polarisation.

Thus, in principle, the 3-dimensional orientation (modulo a mirror
ambiguity) can be derived from the polarisation data only.  However,
these points will need additional investigations before it is
applicable to real data.

\subsection{Shock Properties}

The observed dimensions of a cluster radio relic with a toroidal shape
can be used to get a rough estimate of the shock strength. This is
based on the observation that in the numerical simulations the diameter
$D$ of the spherical cocoon and the major radius of the torus after
the passage of the shock are approximately equal. Since the major and
minor radii of the torus can be read off approximately from a
sensitive high-resolution radio map, the compression of the radio
plasma by the shock can be estimated. In the idealised case of an
initially spherical and finally toroidal radio cocoon, the compression
factor is given by

\begin{equation}
C = \frac{V_{\rm sphere}}{V_{\rm torus}} =  \frac{2\, D^2}{3\,\pi d^2}\,.
\end{equation}
Using Eq. (12), the shock strength can be estimated.  Assuming the
radio cocoon to be in pressure equilibrium with its environment before
and after the shock passage, the pressure jump in the shock is given
by ${P_2}/{P_1} = C^{\gamma_{\rm rp}}$. If the adiabatic index of the
radio plasma $\gamma_{\rm rp}$ is assumed to be known, e.g
$\gamma_{\rm rp}=4/3$ for an ultra-relativistic equation of state, the
shock strength can be estimated. But even if this assumption is not
justified and the geometry deviates from the idealised geometries
assumed here, the strength of the shock wave should be correlated to
the ratio of the global diameter of a toroidal relic and the thickness
of its filaments. Unfortunately, the quality of the best current radio
maps of relics do not yet allow a qualitative comparison of the shock
strength by comparing the $D/d$ ratios of toroidal relics. But these
maps demonstrate that the necessary sensitivity and resolution might
be reached soon.

If, furthermore, the strength of the shock wave of well resolved
cluster radio relics can be estimated independently from X-ray maps of
the IGM, it would be possible to directly measure the adiabatic index
of radio plasma. Even though the present radio and X-ray data do not
have the required accuracy yet, this method will enable us to measure
the unknown equation of state of radio plasma in the future.

In our simulations the adiabatic index of the radio plasma
($\gamma_{\rm rp} =\gamma_{\rm gas} = 5/3$) and the shock strength are
known ($P_2/P_1 = 3.5$ and $P_2/P_1 = 17.4$ for the shock compression
factor $C_{\rm shock}=2$ and $C_{\rm shock}=3.3$ respectively). Thus
we find that the ratio of the length scales $D/d \approx 3$ for
$C_{\rm shock}=2$ and $D/d \approx 5$ for $C_{\rm shock}=3.3$. This is
roughly consistent with our synthetic radio maps. At least the
qualitative correlation of shock strength and diameter ratio is
clearly observed, as can be seen in Figs. \ref{fig:map.WW+70-00} and
\ref{fig:map.SW70-00}. 

\section{Conclusion}

We have presented 3-D MHD simulations of a hot, magnetised bubble that
traverses a shock wave in a much colder and denser environment. This
is assumed to be a fair model for a blob of radio plasma in the IGM
which is passed by a cluster merger shock wave. We have calculated
radio polarisation maps for the relativistic electron population and
computed the spectrum subject to synchrotron-, inverse Compton- and
adiabatic energy losses and gains. These maps show that the shock wave
produces filamentary radio emitting structures and, in many cases,
toroidal structures. Such filaments and tori are indeed observed by
very recent high-resolution radio maps of cluster radio relics (Slee
et al. 2001\nocite{slee2001}). Our simulations find polarisation
patterns which indicate that the magnetic fields are mostly aligned
with the direction of the filaments. This also seems to be the case
for the observed cluster radio relics.

Therefore, we conclude that we have found strong evidence that cluster
radio relics indeed consist of fossil radio plasma that has been
compressed adiabatically by a shock wave, as proposed by En{\ss}lin \&
Gopal-Krishna (2001)\nocite{2001A&A...366...26E}.

The formation of the tori and filaments is not instantaneous. First,
the simulations show a phase in which the radio plasma is strongly
compressed into a flat shape. During this phase a sheet-like radio
relic with a flat spectrum would be observed. Later, the radio plasma
moves towards the edges of this sheet and finally becomes a torus (or
a more complicated, filamentary structure). Since spectral ageing is
likely to have affected these later stages, we expect that on average
the filamentary relics have a steeper, more bent radio spectrum than
the sheet-like ones.

Our simulations indicate that the diameter $D$ of the bubble of radio
plasma remains approximately constant during the passage of the
shock. It was also found that the final structure consists of radio
filaments of small diameter $d$ that are distributed (often in form of
a torus) in a region of size $D$. The compression factor of the radio
plasma is proportional to $(D/d)^2$, and this ratio is a measure of
the shock strength. Thus, the approximate compression factor of the
radio plasma can be read off the radio map.  The local radio
polarisation strongly reflects the complicated magnetic field
structures while the total integrated polarisation of a relic reveals
the 3-dimensional orientation of the shock wave. Since the compression
aligns the fields with the shock plane, the sky-projected field
distribution is aligned with the intersection of the shock plane and
the sky plane. Thus, the direction of the total E-polarisation vector
yields the sky-projected normal of the shock wave. The angle between
the normal of the shock and the plane of the sky can in principle be
estimated from the fractional polarisation of the integrated flux.  We
conclude that this work provides strong evidence that cluster radio
relics are revived bubbles of fossil radio plasma, the so-called {\it
radio ghosts}. Spectral aging arguments (En{\ss}lin \& Gopal-Krishna
2001\nocite{2001A&A...366...26E}) predict the existence of a sizable
population of yet undetected cluster radio relics which are only
observable with sensitive low frequency radio telescopes. More details
of the simulations can be found in En{\ss}lin \& Br\"uggen (2001).

\section*{Acknowledgments}
We thank O.B. Slee, A.L. Roy, M. Murgia, H. Andernach, M. Ehle for
providing us with their observational data prior to publication, and
allowing us to display their 1.4 GHz map of the relic in Abell 85. We
also thank G. Giovannini and L. Feretti for access to their 330 MHz
data of the same relic. We acknowledge A. Kercek's contributions to a
very early stage of this simulation project. Some of the computations
reported here were performed using the UK Astrophysical Fluids
Facility (UKAFF). This work was supported by the European Community
Research and Training Network `The Physics of the Intergalactic
Medium'. TAE and MB acknowledge the award of a European grant to
attend this excellent conference.

\end{document}